# Large-scale defects hidden inside a topological insulator grown onto a 2D substrate

*Danielle Reifsnyder Hickey[1], Ryan J. Wu[1], Joon Sue Lee[2,3], Javad G. Azadani[4], Roberto Grassi[4], Mahendra DC[5], Jian-Ping Wang[4], Tony Low[4], Nitin Samarth[2*], K. Andre Mkhoyan[1*]*

[1]Department of Chemical Engineering and Materials Science, University of Minnesota, Minneapolis, MN 55455

[2]Department of Physics and Materials Research Institute, The Pennsylvania State University, University Park, PA 16802

[3]California NanoSystems Institute, University of California, Santa Barbara, CA 93106

[4]Department of Electrical and Computer Engineering, University of Minnesota, Minneapolis, MN 55455

[5]School of Physics and Astronomy, University of Minnesota, Minneapolis, MN 55455

*Corresponding authors. E-mail: nsamarth@psu.edu (NS) and mkhoyan@umn.edu (KAM)

**Topological insulator (TI) materials are exciting candidates for integration into next-generation memory and logic devices because of their potential for efficient, low-energy-consumption switching of magnetization[1,2]. Specifically, the family of bismuth chalcogenides offers efficient spin-to-charge conversion because of its large spin–orbit coupling and spin–momentum locking of surface states[3,4]. However, a major obstacle to realizing the promise of TIs is the thin-film materials' quality, which lags behind that of epitaxially grown semiconductors. In contrast to the latter systems, the Bi-chalcogenides form by van der Waals epitaxy, which allows them to successfully grow onto substrates with various degrees of lattice mismatch. This flexibility enables the integration of TIs into heterostructures with emerging materials, including two-dimensional materials. However, understanding and controlling local features and defects within the TI films is critical to achieving breakthrough device performance. Here, we report observations and modeling of large-scale structural defects in $(Bi,Sb)_2Te_3$ films grown onto hexagonal BN, highlighting unexpected symmetry-breaking rotations within the films and the coexistence of a second phase along grain boundaries. Using first-principles calculations, we show that these defects could have consequential impacts on the devices that rely on these TI films, and therefore they cannot be ignored.**

The bismuth chalcogenides—specifically $Bi_2Se_3$, $Bi_2Te_3$, and $(Bi,Sb)_2Te_3$—have already demonstrated exciting properties such as topologically protected surface states that can generate spin-transfer torque, undergo current-induced spin polarization, and exhibit ferromagnetic resonance-driven room-temperature spin pumping[2,5–8]. Additionally, this family of materials is at the forefront of integration into devices because its tetradymite crystal structure is two-dimensionally layered, containing covalent bonding within the layers and connected across the basal plane by van der Waals-type interactions[9,10]. This means that, similarly to other 2D materials, the bismuth chalcogenides can grow in an epitaxial-like manner without the stringent requirement for lattice parameter matching that typically accompanies successful epitaxial growth[11–15].

To develop high-quality TI materials, it is critical to understand their atomic structures[3]. Common defects—such as stacking faults, twin boundaries, antiphase boundaries, and strain—have been shown to affect local properties of TIs, from lifting degeneracies to create a net spin current to locally shifting the Dirac states[16–19]. Yet, the more consequential large-scale structural features of these thin films have not been explored. A particularly promising material is $(Bi,Sb)_2Te_3$, in which self-compensation by native defects lowers the bulk carrier density relative to that of other compounds, making it possible to use gating to tune the chemical potential through the Dirac point[4,15]. Here, we used scanning and conventional transmission electron microscopy (S/TEM) to study the structure of $(Bi,Sb)_2Te_3$ thin films grown by molecular beam epitaxy (MBE) onto flakes of the two-dimensional (2D) material hexagonal boron nitride (h-BN), which is a promising dielectric that enables gating of devices[20]. By combining imaging *via* STEM and conventional TEM (CTEM), elemental analysis using energy-dispersive X-ray spectroscopy (EDX) and electron energy-loss spectroscopy (EELS), and *Multislice* simulations, we observed unexpected structural features, such as small rotations within the TI film and a separate phase at grain boundaries. We note that TI films of better crystalline quality can be produced using a different substrate or implementing a different method of growth[21]. However, the fundamental structural features discussed here, such as defects and grain boundaries, are likely to be universal to all TI films.

Devices built from thin films are susceptible to performance variations related to deviations from the single-crystalline structure. To illustrate this, Fig. 1a highlights two different device locations that incorporate TI film regions with dissimilar local structures. The presence of grain boundaries and multiple crystal phases, the stoichiometry of the target phase, and the disruption of a crystal structure's symmetry can all affect how electrons and their spins move across an active region of a device. The idealized crystal model of the $(Bi,Sb)_2Te_3$/h-BN heterostructure investigated here is shown in Fig. 1b. Both the TI and h-BN have layered structures with covalent bonding within either the atomic planes (in h-BN) or five-atom-thick quintuple layers (QLs, in $(Bi,Sb)_2Te_3$), accompanied by van der Waals bonding between the layers. An atomically sharp interface forms between the $(Bi,Sb)_2Te_3$ and h-BN due to the lack of dangling bonds in each material (see Fig. SI1). This is in contrast to TI films grown on other substrates, for which interfacial layers of various thicknesses and compositions have been observed[22–26]. (Bi,Sb)$_2$Te$_3$ has the rhombohedral tetradymite crystal structure R-3m (described throughout this report using hexagonal coordinates), with the covalently bonded QL units stacked three high in the c-direction to complete a unit cell.

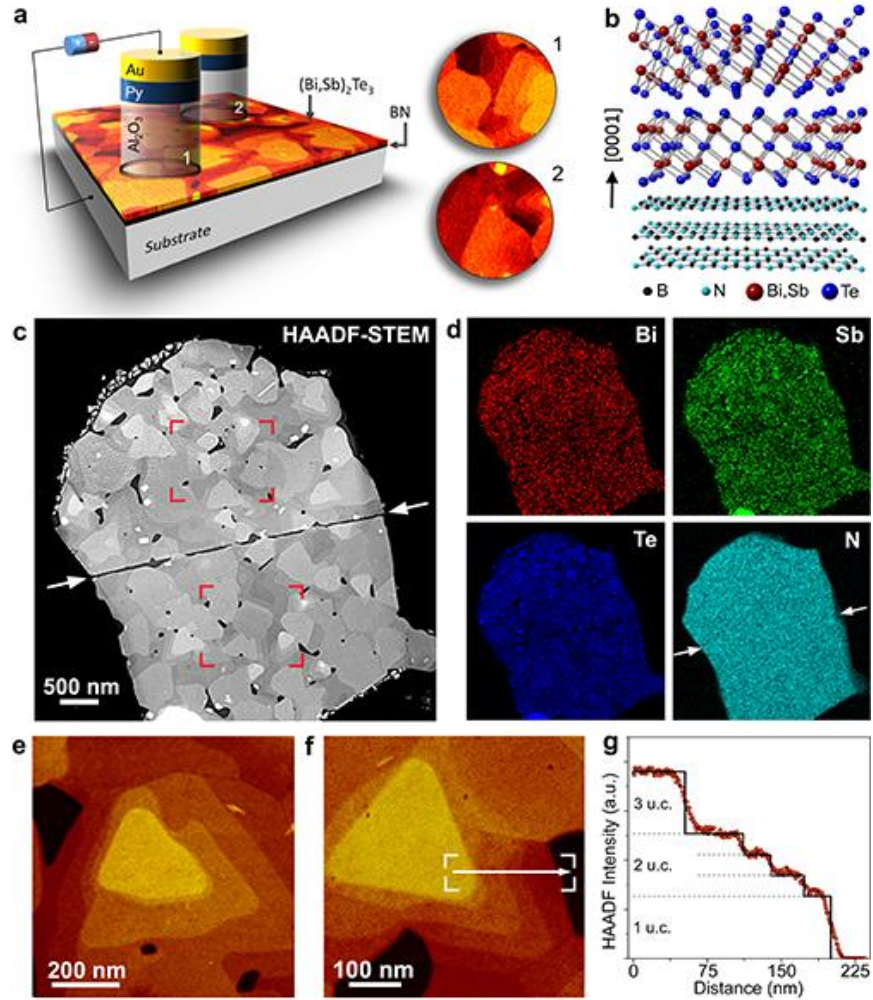

**Figure 1. Overview of (Bi,Sb)$_2$Te$_3$/h-BN heterostructure.** (a) Schematic of devices patterned onto a TI/h-BN heterostructure. The device footprints are indicated by black circles, as shown in top view at right. (b) Cross-sectional model of the TI/h-BN heterostructure. (c) HAADF-STEM image (top view) of the TI film grown on an h-BN flake. (d) The accompanying EDX maps for Bi, Sb, Te, and N. Quantification of two representative film areas indicated by red boxes (one in the center of the region above the break marked with white arrows, and one in the center of the region below) give stoichiometric ratios of 1:1.2:2.6, similar to the expected Bi:Sb:Te ratio of 1:1:3, with slightly more Sb and slightly less Te. (e,f) HAADF-STEM images of examples of triangular features observed in the TI film: (e) an irregular, terraced, triangular feature and (f) a regular, terraced triangular feature. (g) Plot of the HAADF-STEM intensity across the region marked by a white arrow in (f). Black lines are a guide to the eye, showing the steps located at even increments or multiples.

To understand the internal structure of the TI films, this analysis intentionally focuses on data from films that are not completely coalesced. Such films provide better insight into the structural features because they make it possible to directly analyze multiple features associated with grain boundaries and the TI orientation relative to the substrate. In these, and in related, coalesced MBE-grown (Bi,Sb)$_2$Te$_3$ films (see Fig. SI1), grains are several hundred nanometers in size with visible boundaries. The TI films studied here have thickness variations but are approximately two-to-three unit cells thick in the c-direction, atop ~30 layers of exfoliated h-BN, estimated using atomic force microscopy (AFM, not shown). The overall film morphology is shown in the plan-view high-angle annular dark-field scanning TEM (HAADF-STEM) image in Fig. 1c, and its elemental composition is represented by the EDX and EELS elemental maps of the heterostructure presented in Figs. 1d and SI2, respectively. The TI films show many grain-level regular and irregular triangular terraces (Fig. 1e-g), all of which contribute to thickness variations in the TI film. Interestingly, the HAADF-STEM intensity (Fig. 1g), which correlates with sample thickness, decreases across the terraces in 1x and 3x increments, consistent with QL and unit cell (3-QL) steps. This observation suggests that (i) a full unit cell is not necessary for film growth and (ii) grain boundaries in the TI film can follow terracing instead of being sharp vertical boundaries expected according to the crystal symmetry.

Low-magnification CTEM images obtained from the film show striking patterns that provide clues about large-scale features within the film (Fig. 2a,b) that are not visible in HAADF-STEM images or by non-TEM techniques. Entire grains exhibit this feature, and it even persists across boundaries where islands have coalesced (Figs. 2 and SI3). This effect results from small rotations within the TI layers that give rise to several-hundred-nm regions of quasi-periodic Moiré patterns. These patterns can be reproduced by simulations when the two-component (Bi,Sb)$_2$Te$_3$/h-BN heterostructure is divided into a three-layer system: in this case, one layer is h-BN, and the TI is divided into two independent layers, with one TI layer slightly rotated relative to the other (detailed analysis is provided in the SI). The quasi-periodic Moiré patterns observed in the experimental CTEM images can be identified from a series of simulated CTEM images, calculated using the well-established *Multislice* code[27], with different angles for the layer rotations (Fig.

SI4). It should be noted that such quasi-periodic Moiré patterns cannot be reproduced when the heterostructure is treated as a two-layer (TI/h-BN) system: the Moiré patterns in the CTEM images for the two-layer system are highly periodic (see Fig. SI4), as has been observed previously[28,29]. On the other hand, when the TI layer is divided into two segments, the rotations between TI layers within the three-layer structure give rise to the large-period, quasi-periodic features observed in experimental CTEM images.

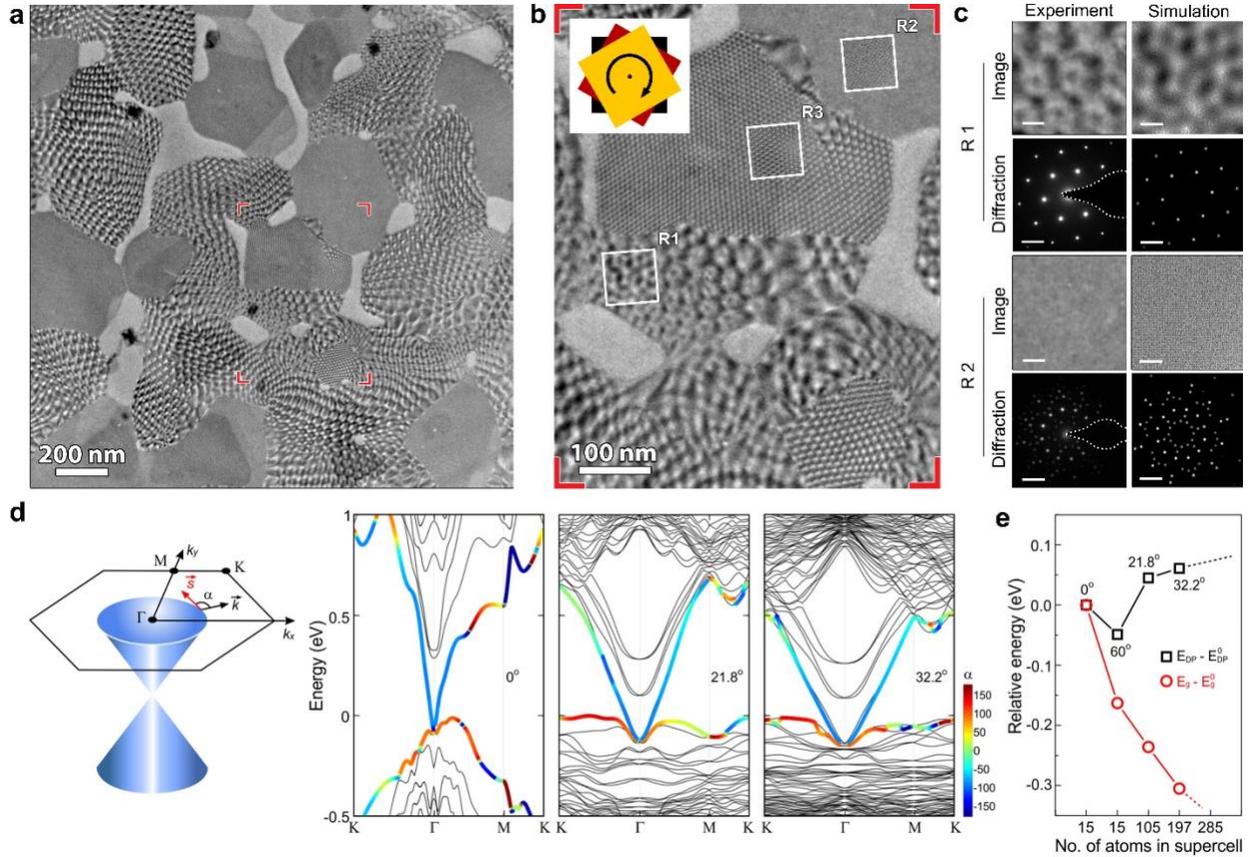

**Figure 2. Rotations within the TI/h-BN heterostructure.** (a) Low-magnification conventional TEM image of the $(Bi,Sb)_2Te_3$/h-BN heterostructure showing large-scale structural features. (b) Magnified section of the image in (a) (indicated by red marks) showing three different types of regions, with *Multislice* simulated images overlaid. Inset: an illustration of the three-layer TI/TI/h-BN model with relative rotations between the layers (black = h-BN, maroon = TI-1, and yellow = TI-2). (c) Experimental and simulated conventional TEM images and diffraction patterns for regions 1 and 2. The position of the beam stop in the experimental diffraction patterns is highlighted with a dotted line. (d) DFT-calculated band structures of three-QL $Bi_2Te_3$ with different amounts of rotation of the top QL: 0°, 21.8°, and 32.2°. The schematic of the simplified Dirac cone with Brillouin zone is shown on the left. In all three band structures, the angle α between wavevector k and spin s of the electron located in the top surface QL is identified by color map. (e) Changes in band gap ($E_g$) and Dirac point ($E_{DP}$) in the band structure as a function of supercell size relative to TI film with no rotation between layers ($E_g^0$, $E_{DP}^0$).

Fig. 2c presents a comparison of experimental and simulated images for the two distinct regions that are observed for this $(Bi,Sb)_2Te_3$/h-BN heterostructure: (R1) regions with large, wavy, quasi-periodic features with a period of tens of nanometers, and (R2) featureless regions that possess no Moiré contrast at low magnification. A third case (R3), which is a subcategory of (R1), exhibits smaller, quasi-periodic features with a period of ~5 nm. The simulated three-layer system with a relative rotation of 0.5° between TI layers is overlaid (region (R1)) onto the experimental CTEM image in Fig. 2b, supporting assignment of the feature as a small rotation of the upper TI layer (Fig. SI4). Region (R3) of the experimental image also contains an overlaid simulated image, which corresponds to the existence of a 2° rotation between TI layers (Fig. SI4). Transitions from small to much larger periodicities are also visible in the experimental data (Fig. 2a,b), indicating that areas exist with both gradual and abrupt changes in relative rotation.

The three cases listed above in fact encompass a wider range of symmetries because of the internal three-fold symmetry of $(Bi,Sb)_2Te_3$ and its ability to form basal twins. These twins can be distinguished in cross section (Fig. SI5), but in the plan-view data presented here, the case of 0° rotation between TI layers resembles that of 60° (basal twin) and 120° (identical to 0° due to three-fold rotation). In each case, a small rotation results in Moiré patterns such as those observed in Fig. 2a,b (i.e., 0° and 60° are similar, and 2°, 58°, and 62° resemble each other). A thorough series of simulations (Fig. SI6), however, shows that the Moiré patterns also disappear at intervals of 30°. Here, electron diffraction patterns can distinguish between 30° increments. Fig. 2c shows the experimental and simulated electron diffraction patterns corresponding to regions (R1) and (R3). The latter case (30°-rotated TI) corresponds to a crystallographic epitaxial relationship between $(Bi,Sb)_2Te_3$ and h-BN defined by $(Bi,Sb)_2Te_3$ (-2110) || h-BN (1-100) and $(Bi,Sb)_2Te_3$ (0001) || h-BN (0001). The diffraction patterns in Fig. 2c confirm that the grain with featureless contrast in fact is due to a 30° rotation of the TI relative to the h-BN, not a multiple of 60° rotation. The fact that simulations of small rotations within the TI layer at 30° relative to the h-BN substrate do not show large-scale Moiré fringes (Fig. SI6) suggests that small rotations within the TI layer can be identified here from CTEM imaging for the 0° and 60° cases but not for the 30° case.

Using density functional theory (DFT), we investigate the effect of rotations on the spin–momentum locking of pure $Bi_2Te_3$ that is one unit cell thick (3 QLs), chosen as a model system (see Fig. SI7). In the case of perfect spin–momentum locking, the angle between the projection of the in-plane spin vector and the wavevector $\vec{k}$ is $\alpha = 90°$. For the pristine TI, due to time-reversal and inversion symmetry, the top and bottom surface bands are degenerate (Fig. 2d)[3]. When one QL is rotated and the inversion symmetry is broken in the TI, the degeneracy of the surface bands is lifted (Fig. 2d). Interestingly, the results show that 90° spin–momentum locking in the Dirac bands in the vicinity of the Γ point is mostly preserved even in the rotated TI, although deviation from perfect 90° can be seen, especially away from the Γ point. This is shown in Fig. 2d, where the values of α are plotted for the top surface (for the bottom surface, see Fig. SI8). However, the Dirac point is lowered with respect to the Fermi energy as the rotation angle decreases (or as the size of the supercell increases). Even more importantly, the band gap of the TI film also decreases with the size of the supercell (Fig. 2e). Such a dramatic reduction of the band gap was unexpected, and it will have a considerable effect on device performance. It should be noted that for 3-QL-thick $Bi_2Te_3$ film the band gap is larger than that for bulk material due to quantum confinement (see Fig. SI9). Calculations also show that the rotation-induced changes to the band structure, including the spin–momentum locking distribution, are dramatically different than those due to quantum confinement.

In addition to rotations between TI layers in these films, some grain boundaries harbor a second phase. This phase, shown at a grain boundary in Fig. 3a and at higher magnification in Fig. 3b, is the tellurium phase with space group $P3_121$ (see phase diagram in Fig. SI10)[30]. In plan view, it is visible in atomic-resolution HAADF-STEM images. Its characteristic triplet of atoms is distinct from the hexagonal arrangement of atoms in the $(Bi,Sb)_2Te_3$ film. An EDX line scan across the grain boundary, shown in Fig. 3c, confirms its predominantly Te composition with < 20 at% Bi and Sb in this region. While it is unexpected to observe such Te-phase grain boundaries in otherwise nearly stoichiometric MBE-grown $(Bi,Sb)_2Te_3$ films, the coexistence of $Bi_2Te_3$ and Te phases has been observed previously, when Te was used as a seed for the growth of $Bi_2Te_3$ nanoplates[31,32], and the two phases are expected to coexist at Te at% greater than 60 (Fig. SI10). Interestingly, TI regions adjacent to these grain boundaries show enhanced Te depletion in the $(Bi,Sb)_2Te_3$ film, with stoichiometry of 1:1:1.6 (Fig. 3c) instead of the value of 1:1.2:2.6 measured across multiple grains (Fig. 1). This suggests that diffusion of elements from neighboring grains could be the origin of this second phase at grain boundaries. The Te phase does not appear to form preferentially along a specific type of grain boundary. The example in Fig. 3a occurs at a 30° grain boundary, but the phase also occurs at a triple grain boundary, as shown in Fig. 3d, which has grains with relative orientations of 10°, 20°, and 30°. In some cases, the Te phase occurs in pockets, and it persists for several unit cells into the grain. In other cases, it is only as wide as one or a few triangular clusters of atoms (see Figs. 3 and SI10 for examples). Regions of the Te-phase along grain boundaries in some cases are as long as 100 nm (Fig. 3e) and could span an entire device. As these grain boundaries are electrically p-type semiconducting[31], their impact on device performance can be consequential.

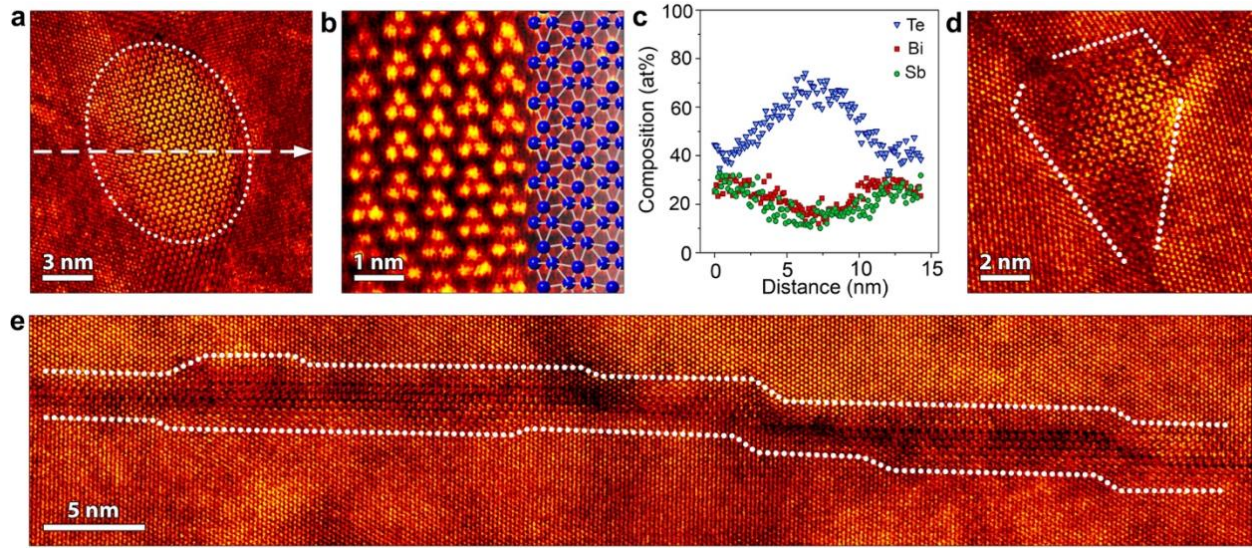

**Figure 3. Grain boundaries with Te phase.** (a) HAADF-STEM image of a Te pocket at a grain boundary. (b) Atomic-resolution HAADF-STEM image showing the atomic structure of the Te phase with a ball-and-stick model overlay. (c) EDX line scan across the grain boundary, showing composition vs. distance, obtained along the arrow shown in (a). The dashed line in (a) shows the direction of the line scan collected. (d) HAADF-STEM image of the Te phase at a triple grain boundary. (e) High-resolution HAADF-STEM image of a large-scale grain boundary that contains the Te phase. Notably, the Te phase exists on both sides of the grain boundary, although only on one side at any given position along the grain boundary.

However, not all grain boundaries contain the Te phase. The MBE growth process is known to create islands of TI that coalesce during growth[21], and a closer look at the grain boundaries of coalesced islands with atomic-resolution HAADF-STEM imaging reveals that grains often grow together with 0°, 2°, 7°, and 30° misorientations (Fig. 4). As expected, the 2° and 7° grain boundaries actually each consist of an array of dislocations (Figs. 4b,d and SI11), as has been observed in thicker $Bi_2Se_3$ films[19]. The 0° and 30° orientations are very common in the film, and small-angle dislocation boundaries are observed less frequently. Due to the rhombohedral R-3m crystal structure of $(Bi,Sb)_2Te_3$, the apparent 0° boundary between TI islands has indistinguishable symmetry in the FFT from that of a 60° grain boundary. In fact, the presence of 60° rotated triangular regions can be seen in low-magnification STEM images (Fig. SI1).

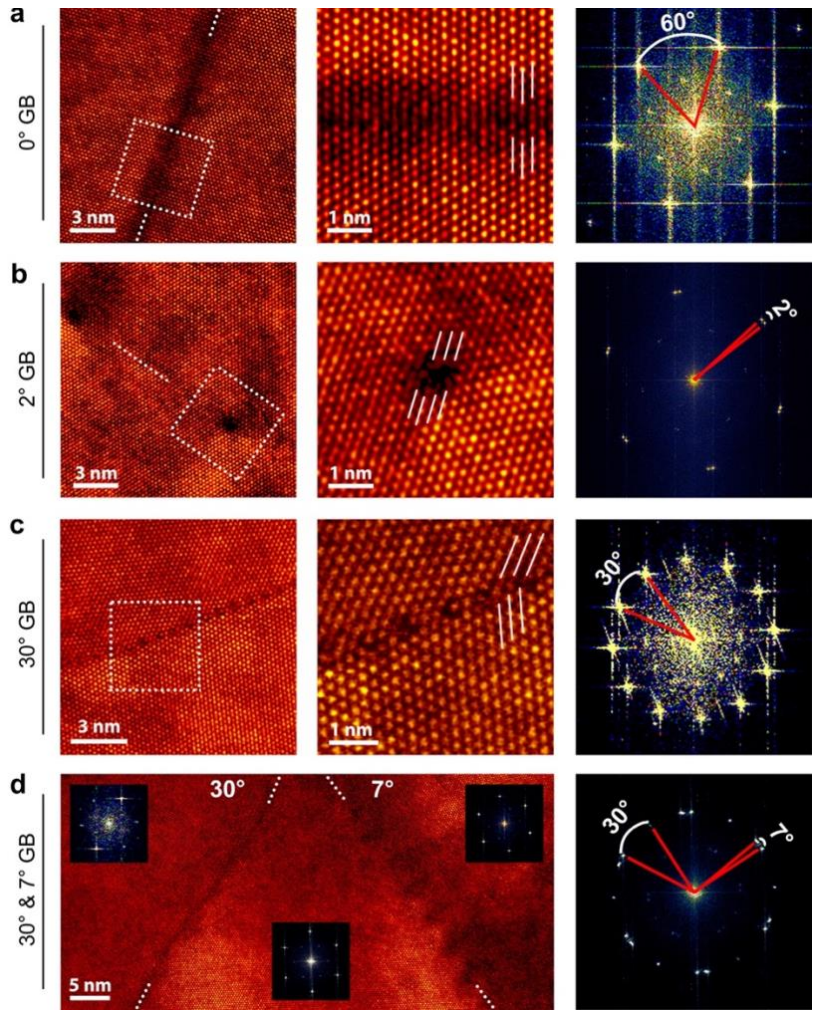

**Figure 4. Atomic-resolution HAADF-STEM imaging of grain boundaries.** (a) Example of a boundary with no misorientation, shown at lower (left) and higher (center) magnification, along with an FFT from the two grains (right). 0° and 60° rotations between grains are indistinguishable. (b,c) Examples of a 2° dislocation boundary and a 30° grain boundary, respectively, shown at lower (left) and higher (center) magnification, along with FFTs from the grains (right). Dashed boxes indicate the regions selected for presentation at successively higher magnification. (d) Example of three grains connected by 30° and 7° grain boundaries. Insets are the corresponding single-grain FFTs. FFT across the three grains, with the 30° and 7° angles labeled (right).

In conclusion, we have discovered large-scale, dominant features of TI films in emerging heterostructures that cannot be readily identified by other techniques, including internal TI film rotations and coexistence with a Te phase along grain boundaries. Although the details of films vary with growth conditions and the selection of the substrate, such features as film morphology, variations in stoichiometry that can lead to secondary phases, and symmetry breaking between film layers can have application-limiting effects for devices. In TIs, especially, for which exciting properties emerge in the thin-film geometry, it is critical to consider the influence of these persistent large-scale, sometimes micron-size, structural variations that could limit the ultimate device performance.

## Methods

*Heterostructure formation*

Heterostructures of the topological insulator $(Bi,Sb)_2Te_3$ on hexagonal boron nitride (h-BN) were formed using mechanical exfoliation of h-BN and molecular beam epitaxy (MBE) growth of $(Bi,Sb)_2Te_3$. The Scotch tape method was used in air to exfoliate h-BN flakes, which were then deposited onto a $SiO_2$(300 nm)/Si wafer. After being transferred into the ultrahigh vacuum MBE chamber, this $Si/SiO_2$/h-BN substrate was outgassed prior to TI growth. Then, two-step growth of $(Bi,Sb)_2Te_3$ was performed. The first two quintuple layers were grown at a lower temperature (150-175°C), after which the rest of the film was grown at higher temperature (290-320°C).

*TEM measurements*

Conventional and scanning transmission electron microscopy were performed on plan-view samples of the $(Bi,Sb)_2Te_3$/h-BN heterostructure that had been exfoliated from the $SiO_2$/Si substrate. The flakes were transferred by a dry, polymer-based method using a polydimethylsiloxane (PDMS) stamp coated with poly(propylene carbonate) (PPC). Prior to TEM analysis, the PPC coating was dissolved using anisole. Additional cross-sectional $(Bi,Sb)_2Te_3$ samples were prepared using an FEI Quanta 200 3D dual-beam $Ga^+$ focused ion beam (FIB) at 30 and 10 kV. Conventional TEM measurements were performed on an FEI Tecnai G2 F30 TEM at an accelerating voltage of 300 kV. High-angle annular dark-field scanning TEM (HAADF-STEM) imaging, energy-dispersive X-ray spectroscopy (EDX) and electron energy-loss spectroscopy (EELS) were performed at 200 kV on an FEI Titan G2 60-300 aberration-corrected S/TEM equipped with a Schottky X-FEG gun. The electron probe convergence angle was 16 mrad, the probe current was ~ 200 pA, the inner angle of the HAADF detector was 53.9 mrad, and the EELS collection angle was 14.3 mrad. Energy-dispersive X-ray spectroscopy (EDX) was performed using a Super-X quad-SDD windowless in-polepiece EDX detector, and EELS was performed using a Gatan Enfinium ER spectrometer.

*Image simulations*

CTEM images and diffraction patterns were simulated using the *Multislice* method[33], implemented using the TEMSIM code developed by Kirkland (2010)[27]. All TEM simulations were performed using a 300 kV electron beam with Cs3 = 1 mm, Cs5 = 0 mm, and defocus $\Delta f$ = 0 Å to approximate experimental TEM operating conditions. Lattice parameters used for the simulated $(Bi,Sb)_2Te_3$/h-BN heterostructure were a = 4.33 Å and c = 30.4 Å[34] for $(Bi,Sb)_2Te_3$ and a = 2.504 Å and c = 6.660 Å[35] for h-BN, unless otherwise specified. The $(Bi,Sb)_2Te_3$ unit cell and h-BN interlayer spacings were 2.665Å[34] and 3.3Å,[35] respectively. The distance separating $(Bi,Sb)_2Te_3$ from h-BN was set to 3 Å. Effects of thermal displacements were simulated by averaging 20 frozen phonon configurations at 300 K for each image and diffraction pattern. Root mean square thermal displacement values used were 0.110 Å (B), 0.096 Å (N),[36] 0.087 Å (Bi), 0.087 Å (Sb), and 0.100 Å (Te)[37].

Diffraction pattern simulations used a $(Bi,Sb)_2Te_3$/h-BN supercell of 248 x 238 $A^2$ with a thickness defined by 2 unit cells of $(Bi,Sb)_2Te_3$ and 30 atomic layers of h-BN. For CTEM simulations, the supercell area was quadrupled in order to capture long-range features observed in experimental images. The heterostructure was built from three-dimensional replication of unit cell coordinates, and a rotation matrix was employed to independently rotate the h-BN and two TI layers, making it possible to explore the full range of rotation and the effects of treating the structure as either a two- or three-layer system. The probe and transmission functions were calculated with 8192 x 8192 pixelation. Slice thickness was set to 1 Å. For diffraction simulations, a slightly converged probe $\alpha$ = 0.1 mrad was used to smear the effects of pixelation.

*Data processing*

All atomic-resolution STEM images were low-pass filtered to exclude information below 1 Å, the measured experimental STEM resolution, in order to improve the signal-to-noise ratio. The fast Fourier transform (FFT) of a HAADF-STEM image of a standard Au sample was used to determine the STEM spatial resolution. Simulated CTEM images were first convoluted with a two-dimensional Gaussian function with a FWHM of ~ 2 Å to mimic experimental CTEM image resolution. Subsequently, the convoluted images were cropped in the center to 650 Å by 650 Å images with renormalized intensities. Likewise, simulated and experimental diffraction patterns were convoluted using a two-dimensional Gaussian function with a FWHM of ~0.05 Å$^{-1}$ to enhance the visibility of the diffraction spots.

*DFT calculations*

Spin-polarized DFT-based calculations were performed using the VASP package[38]. Projector-augmented wave potentials (PAW)[39] and the exchange correlation approximated by the Perdew–Burke–Ernzerhof (PBE) functional[40] were employed. A plane-wave basis set with kinetic energy cutoff value of 300 eV was used for all calculations. Integration over the 2D Brillouin zone (BZ) was performed using a (20×20×1), (11×11×1), and (6×6×1) Γ-centered Monkhorst-Pack *k*-points mesh for the 0°, 60°, 21.8°, and 32.2° rotated TIs, respectively. Only one of the three QLs in this one-unit-cell-thick TI was rotated. The break condition for the electronic self-consistent loop was chosen to be 10$^{-5}$ eV. To calculate spin densities, wave functions using the WaveTrans code[41] were employed, followed by projecting spin expectation values onto the top TI layer. We used a vacuum size of 33 Å along the z-direction between supercells.

**Acknowledgements**

This work was supported by C-SPIN, one of six STARnet program research centers. This work utilized (1) the College of Science and Engineering (CSE) Characterization Facility, University of Minnesota (UMN), supported in part by the NSF through the UMN MRSEC program (No. DMR-1420013); and (2) the CSE Minnesota Nano Center, UMN, supported in part by NSF through the NNIN program. The authors would also like to thank Mr. Prashant Kumar and Dr. Jong Seok Jeong for assistance with figure illustrations.

**Contributions**

D.R.H. performed the TEM experiments and *Multislice* simulations, analyzed the data, and wrote the manuscript. R.J.W. contributed to data processing, figure preparation, and building the TI/h-BN rotated supercell for *Multislice* simulations. J.S.L. grew the films. J.G.A., R.G. and T.L. performed the DFT calculations. M.DC and J.-P.W. contributed into understanding of the results. K.A.M. designed the experiments and supervised the analysis, along with N.S., who also supervised film growth. All authors discussed the results and contributed to the manuscript.